\documentclass[twocolumn,numberedappendix]{emulateapj}
\usepackage{graphicx}
\usepackage{hyperref}
\usepackage{amssymb}
\usepackage{amsmath}
\usepackage{units}
\usepackage{fancyvrb}
\newcommand{\figwidth}{3.25in}

\newcommand{\twopartdef}[4]
{
	\left\{
		\begin{array}{ll}
			#1 & \mbox{if } #2 \\
			#3 & \mbox{if } #4
		\end{array}
	\right.
}

\newcommand{\comments}[1]{}

\newcommand{\diver}[1]{\nabla \cdot \mathbf{#1}}

\def\app#1#2{%
  \mathrel{%
    \setbox0=\hbox{$#1\sim$}%
    \setbox2=\hbox{%
      \rlap{\hbox{$#1\propto$}}%
      \lower1.1\ht0\box0%
    }%
    \raise0.25\ht2\box2%
  }%
}

\DefineShortVerb{\|}

\begin{document}

\title{The Efficiency of Second-Order Fermi Acceleration by Weakly
  Compressible MHD  Turbulence} 
\author{Jacob W. Lynn\altaffilmark{1,2}, Eliot Quataert\altaffilmark{2}, Benjamin D. G. 
  Chandran\altaffilmark{3}, \& Ian J. Parrish\altaffilmark{2,4}
  } 

\altaffiltext{1}{Physics Department, University of California,
  Berkeley, CA 94720; jacob.lynn@berkeley.edu} 
\altaffiltext{2}{Astronomy Department and Theoretical Astrophysics 
  Center, University of California, Berkeley, CA 94720} 
\altaffiltext{3}{Space Science Center and Department of Physics,
  University of New Hampshire, Durham, NH 03824}
\altaffiltext{4} {Present address: Canadian Institute for Theoretical
    Astrophysics, 60 St. George Street, University of Toronto,
  Toronto, ON M$5$S $3$H$8$, Canada}

\begin{abstract}
We investigate the effects of pitch-angle scattering on the
efficiency of particle heating and acceleration by MHD turbulence
using phenomenological estimates and 
simulations of non-relativistic test particles interacting with strong, subsonic MHD turbulence.  We  include an imposed pitch-angle scattering rate, which is meant to approximate the effects of high frequency plasma waves and/or velocity space instabilities.  We focus on plasma parameters similar to those found in the near-Earth solar wind, though most of our results are more broadly applicable.
An important control parameter is the size of the particle mean free
path $\lambda_{\mathrm{mfp}}$ relative to the scale of the turbulent
fluctuations $L$.   For small scattering rates, particles  interact  quasi-resonantly
with turbulent fluctuations in magnetic field strength. Scattering
increases the long-term efficiency of this resonant heating by factors
of a few-10, but the distribution function does not develop a significant non-thermal power-law tail.
For higher scattering rates,
the interaction between  particles and turbulent fluctuations becomes non-resonant, governed by  particles  heating and cooling adiabatically as they encounter turbulent density fluctuations.
Rapid pitch-angle scattering can produce a power-law tail in the proton distribution function but this requires  fine-tuning of parameters. Moreover, in the near-Earth solar wind, a significant power-law tail cannot develop by this mechanism  because the particle acceleration timescales are  longer than the adiabatic cooling timescale set by the expansion of the solar wind.    Our results thus imply that MHD-scale turbulent fluctuations are unlikely to be the origin of the $v^{-5}$ tail in the proton distribution function observed in the solar wind. 

\

\end{abstract}


\section{Introduction}
\label{sec:introduction}

Given the ubiquity of MHD turbulence in space and astrophysical plasmas, it is critical to understand the efficiency of particle heating and acceleration by MHD-scale turbulent fluctuations.   In this paper we address this problem using numerical simulations of test particles interacting with strong MHD turbulence.

According to quasi-linear theory, high-gyrofrequency particles interacting with low-frequency, long-wavelength
turbulence in collisionless plasmas do so primarily through resonant
interactions with compressible modes \citep{Achterberg1981}. For low amplitude turbulent fluctuations, conservation of magnetic moment ensures that changes in perpendicular velocity are reversible,
and so most acceleration is in the parallel direction, through the
mirror force at $\beta \gtrsim 1$ and parallel electric fields at $\beta \lesssim 1$. Quasilinear theory predicts that only particles with
precisely the wave velocity experience significant acceleration over
long times \citep{Kennel1966,Jokipii1966}. However, because
waves are not long-lived 
in realistic MHD turbulence, the resonance is significantly broadened, and many particles
can experience  acceleration \citep{Bieber1994a,Shalchi2004,Yan2008a,Lynn2012}.

As a result of this parallel diffusion, an initially isotropic distribution 
function becomes anisotropic over time, with  more energy in the $v_\parallel$
direction.    However, as high-velocity particles are continually 
accelerated, the heating and acceleration of a distribution of particles becomes progressively less effective. This is because the heating efficiency is substantially lower for particles
with velocities much larger than the resonant wave velocity.   Moreover, the mirror force depends on $v_\perp$ but not $v_\parallel$.  

In the solar wind and other astrophysical plasmas,
small-scale microinstabilities such as the mirror, firehose, and ion cyclotron
instabilities can become important if the distribution function
becomes too anisotropic \citep{Kasper2002,Sharma2007,Schekochihin2008a,Bale2009a}. These
instabilities will tend to quickly 
isotropize distributions functions upon their onset \citep{Hellinger2008}.  A population of
high frequency ion-cyclotron or whistler waves excited independent of velocity space instabilities can have the same effect for super-thermal particles. 

If pitch-angle scattering is effective,  particles which gain parallel velocity
 are efficiently isotropized, increasing the efficiency of particle heating and acceleration.   If scattering rates are even higher, such that the mean free
paths are smaller than the size of the turbulent eddies, then resonant wave-particle interactions no longer occur because scattering destroys the wave-particle phase coherence.  Instead, particles
are  locked into the fluid motions and heat and cool
adiabatically along with compressions and expansions of the
fluid \citep{Skilling1971}. Particles  then gain or lose energy
irreversibly by spatially diffusing to neighboring uncorrelated
eddies. This non-resonant process can contribute significantly to the overall heating 
rate.   It has also been proposed as an
efficient particle acceleration mechanism that can generate a  population of non-thermal, high energy particles \citep{Ptuskin1988a}.  Recently, \citet{Fisk2008} argued that  interactions of this kind with compressional turbulence could produce the $v^{-5}$ tail in the proton distribution function observed at energies $\sim 0.01-1$ MeV in the solar wind and outer heliosphere.

In this paper, we determine the efficiency of particle heating and acceleration in MHD turbulence by implementing  pitch-angle scattering of test particles evolving in numerical simulations of strong MHD turbulence; the scattering mimics the isotropization caused by plasma
waves and/or instabilities. There is a substantial literature
investigating the interaction of test particles with an ensemble of
linear waves, intended to represent turbulence
\citep[e.g.][]{Giacalone1994,Mace2000,Qin2012,Tautz2013}.  Our
numerical results relax many of the simplifying assumptions of these
simulations and of standard quasi-linear theory and thus provide a
more robust estimate of particle energization by MHD-scale turbulent
fluctuations in heliospheric and astrophysical plasmas.   In our
calculations, we restrict our analysis to non-relativistic particles
and focus on low-frequency, weakly-compressible MHD  turbulence that
consists primarily of 
Alfv\'{e}n and slow waves, with a small fraction of the energy in fast waves.  This is qualitatively consistent with  MHD turbulence in the
solar wind (e.g., \citealt{Yao2011,Howes2012}).  

In \S \ref{sec:transportProperties}, we summarize phenomenological
arguments from the literature on the
 heating and acceleration of test particles  in the presence of turbulence and
 pitch-angle scattering.  These analytic estimates provide
valuable context for interpreting our  test particle simulations. We
describe our numerical methods in \S
\ref{sec:numericalMethods}, and our test particle simulation results in \S \ref{sec:particleTransport}. In \S \ref{sec:powerLaw} we discuss whether velocity diffusion in the presence of pitch-angle scattering can produce power-law tails in the distribution function.   We then summarize our results and discuss their implications (\S  \ref{sec:conclusions}).

\section{Velocity diffusion:  Analytic Estimates}
\label{sec:transportProperties}

The interaction of high-gyrofrequency test particles with
low-frequency MHD turbulence leads to primarily parallel velocity
diffusion.  Perpendicular velocity diffusion is largely suppressed
due to the approximate conservation of $\mu=v_\perp^2/B$ (see \citealt{Chandran2010} for a detailed discussion of the conditions under which $\mu$ conservation is violated). Efficient
pitch-angle 
scattering converts this parallel velocity diffusion into isotropic velocity
diffusion, described by
\begin{equation}
  \label{eq:isotropicDiffusion}
  \frac{\partial f}{\partial t} = \frac{1}{v^2}
  \frac{\partial}{\partial v} \left( v^2 D \frac{\partial f}{\partial
      v}\right), 
\end{equation}
where $f(v)$ is the 3D distribution function, normalized by $n_0 =
\int d^3 \mathbf{v} f(\mathrm{v})$ \citep{Achterberg1981}.

In this paper we focus on cases in which the pitch-angle scattering rate $\nu$ is sufficiently high
that the distribution function remains isotropic. The timescale over
which 
the distribution function at velocity $v$ would become significantly anisotropic in the case of primarily parallel velocity diffusion can be approximated by
$\tau_{\mathrm{a}} \sim 
v^2/D(v)$.
Thus,  the distribution function will be roughly isotropic at velocity $v$ if $\nu \gg D/v^2$.   In what follows we present results without an explicit velocity dependence of the pitch-angle scattering rate $\nu$.   However, it is quite plausible that in reality the pitch-angle scattering rate does depend on velocity, and, in particular, that thermal and super-thermal particles have different scattering rates.    Because our numerical results in \S \ref{sec:particleTransport} are based on test particle calculations, the case of a velocity-dependent scattering rate can be obtained by interpolating between our numerically determined velocity diffusion coefficients for different values of $\nu$.

Analytic treatments of the heating and acceleration of particles by  MHD turbulence have considered primarily  two distinct  limits.   We briefly review these results  as they provide important context for interpreting  our numerical simulations.

In the low scattering rate limit, the
underlying interactions are essentially still resonant, as in a collisionless plasma. In order for
an interaction with fluctuations on a scale $l$ to be resonant, the
mean free path of a particle due to scattering must satisfy $\lambda_{\rm mfp} \gg l$, where  
 the mean free path is given by $\lambda_{\rm mfp} \simeq v / \nu$. This nearly resonant case is 
discussed in \S \ref{sec:resonantDiffusion}. 

The second limit is when the particles are effectively collisional due to the scattering, i.e., $\lambda_{\rm mfp} \ll l$.  Particles
 now heat and cool
adiabatically along with density fluctuations in the
fluid.  This limit is   discussed in more detail in \S \ref{sec:ptuskinA}. 

\subsection{Resonant velocity diffusion}
\label{sec:resonantDiffusion}

For low scattering rates, the interactions between turbulent fluctuations and particles are similar to those in the collisionless limit, but the scattering maintains a nearly isotropic distribution function.  
The isotropic diffusion coefficient $D$ is given by an appropriate average of the parallel velocity diffusion coefficient
over the pitch-angle cosine $\alpha = v_\parallel / v$, 
\begin{equation}
  \label{eq:diffusionCoefficient}
  D = \frac{1}{2}\int_{-1}^1 d\alpha \,\alpha^2 D_\parallel,
\end{equation}
where $D_\parallel$ is the parallel diffusion coefficient in the
absence of scattering. We assume that $D$ and $f$ will only depend on
the isotropic speed, $v = \vert \mathbf{v} \vert$. 

In the low scattering rate limit, the key interactions that set $D_\parallel$ (and hence the isotropic diffusion coefficient) are approximate resonances between magnetic compressions and particles (also sometimes known as transit time damping, TTD).   In quasi-linear theory, these magnetic compressions are due to fast or slow magneto-sonic waves and the resonances show up as strict delta functions of the form $D_\parallel \propto \delta(v_\parallel - v_p)$ where $v_p$ is the phase speed of the wave with the largest magnetic compressions.     In the case of strong MHD turbulence, however, quasilinear theory is quantitatively inapplicable for describing the wave-particle interactions, due to the
short  lifetimes of waves in turbulence and to the nonlinearity of
the interaction. This has motivated a number of 
phenomenological models for resonance-broadened interactions between
particles and  fluctuations in MHD turbulence. In \cite{Lynn2012} we calibrated analytic
expressions for $D_\parallel$ including resonance broadening  against test particle simulations.   Among other things, we showed that  the exponential time decorrelation of waves often assumed in the literature, which leads to a Lorentzian broadening of the wave-particle resonance, is not consistent with  test particle simulations; instead a much more rapid Gaussian time decorrelation is required.

We parameterize the resulting diffusion coefficient as
\begin{equation}
  \label{eq:TTDParameterized}
  D_\parallel \propto v_\perp^4 \twopartdef{1}
  {\vert v_\parallel \vert < v_p} { \vert \frac{v_p}
    {v_\parallel} \vert ^{\beta}} {\vert v_\parallel \vert >v_p}
\end{equation}
where $v_p$ is the phase velocity of the magnetic compressions (fast or slow waves in quasilinear theory) and $\beta=3$ for
a resonance-broadening model with a Gaussian time decorrelation for the waves comprising the turbulence.  The functional form in equation \ref{eq:TTDParameterized} is quite different 
from the $D_\parallel \propto \delta(v_\parallel - v_p)$ of TTD in
linear theory; resonance broadening allows many more
particles to interact with the turbulent fluctuations. Using this
simplified form for the diffusion coefficient, we perform 
the integral in equation \ref{eq:diffusionCoefficient} to estimate the isotropic velocity diffusion coefficient  in the low scattering rate limit:
\begin{equation}
  \label{eq:simpleD}
  D \propto \twopartdef {8 v^4 / 105} {v < v_p}
  {v_p^3 \left( v \log{(\frac{v} {v_p}}) - \frac{5v} {12} + \frac{3
    v_p^2} {5 v} - \frac{3 v_p^4} {28 v^3} \right)} {v > v_p},
\end{equation}
For high velocities, equation \ref{eq:simpleD} predicts $D \propto v
\log{v}$.

In addition to the quasi-resonant diffusion described by equation
\ref{eq:simpleD}, there is a contribution from non-resonant
interactions with large-scale eddies. These interactions consist of particles being flung along moving, curved
magnetic field lines, like beads on a wire, and are most important for
low velocity (sub-thermal) particles.    We term these Fermi Type-B (FTB)
interactions.   Diffusion coefficient in this
limit are calibrated against test particle simulations in \citet{Lynn2012}.

\subsection{Non-Resonant Velocity Diffusion}
\label{sec:ptuskinA}

If the scattering rate is sufficiently
high that the mean  free path of a particle becomes smaller than the
size of an eddy (i.e., $\lambda_{\mathrm{mfp}} \ll l$), the particle acceleration mechanism becomes 
entirely non-resonant and the results of the previous section do not apply. Instead, particles become approximately ``locked'' into the scattering frame and undergo
adiabatic increases (decreases) in 
energy as the large scale eddies converge (diverge); spatial diffusion out of the compressible fluctuations makes this energy change diffusive rather than adiabatic.   Note that in this case it is the turbulent density fluctuations rather than the magnetic field compressions that  the particles tap into.  In particular, the important density changes are those associated with compressions and rarefactions of the fluid, as in sound waves, not density jumps due to  pressure-balanced entropy modes.   Quantitatively,
this energy change may be expressed as
\begin{equation}
  \label{eq:skillingMomentum}
  \frac{dy}{dt} = -\frac{\nabla \cdot \mathbf{u}}{3},
\end{equation}
where $\mathbf{u}$ is the fluid velocity, and we have introduced the
variable $y \equiv \ln{\left(p/p_0 \right)}$, where $p_0$ is the
initial momentum of a particle \citep{Skilling1975}.

\citet{Ptuskin1988a} derived the velocity diffusion coefficient
for these non-resonant interactions assuming isotropic spatial diffusion and that the scatterers are at rest relative to the ambient fluid (see also, e.g., \citealt{Chandran2003a} and \citealt{Jokipii2010} for related studies).   The latter assumption is plausibly appropriate if non-propagating instabilities such as the firehose and mirror instabilities dominate pitch-angle scattering but not necessarily if (say) Alfv\'en waves do.   We consider moving scatterers in \S \ref{sec:moving}.

Although spatial diffusion in MHD is primarily along magnetic fields lines (and thus anisotropic), Ptuskin's calculation with isotropic diffusion captures many of the key results found in the more general MHD calculation. The results are simplest 
in the limit that particles diffuse across the largest scale
eddies before the crossing time of compressible waves,
$\tau_{\mathrm{diff}} 
\ll \tau_w$, i.e., $v^2 \gg v_w l_D \nu$, where $v_w \simeq c_s$ is
the phase velocity of compressible waves, $l_D$ is the outer-scale of the turbulent fluctuations, and $\tau_w  = l_D/v_w$.
A simplified  phenomenological version of Ptuskin's
derivation, as applied to the outer-scale eddies, is as follows. The correlation time of 
the particle-turbulence interaction is the time
required to cross one eddy, because neighboring eddies are
uncorrelated: $\tau_{\mathrm{corr}} \sim \tau_{\mathrm{diff}} \sim
l_D^2/(2 \kappa)$, where $\kappa$ is the spatial diffusion 
coefficient. The rms change in $y$ as a particle diffuses across 
an eddy is given from Equation \ref{eq:skillingMomentum} by $\delta y
\sim - \tau_{\mathrm{diff}} \left(\nabla \cdot
  \mathbf{u}\right)_{\mathrm{D}}/3$, where $\langle \nabla \cdot
\mathbf{u}\rangle_{\rm D}$ is the rms velocity divergence associated
with the outer-scale (driving-scale) eddies.   Thus, the resulting velocity 
diffusion coefficient is
\begin{equation}
  \label{eq:ptuskin_D}
  D \sim v^2 \frac{\langle \delta y^2\rangle}{2 \tau_{\mathrm{diff}}}
  \sim v^2 \frac{l_D^2 \left(\nabla \cdot
  \mathbf{u}\right)_{\mathrm{D}}^2}{36 \, \kappa} \ \ \ (\tau_{\rm diff} \ll \tau_w),
\end{equation}
which is very similar to equation 21 of \citet{Ptuskin1988a}, up to
dimensionless coefficients and the identification of $\left(\nabla \cdot
  \mathbf{u}\right)_{\mathrm{D}}$ as a proportionality constant
times $u_{\mathrm{rms}} / l_D$.\footnote{Ptuskin's notation 
for spatial and velocity diffusion coefficients is the opposite of ours
(we label them as $\kappa$ and $D$, respectively).}   Note that for a fixed pitch-angle scattering rate $\nu$, $\kappa \sim v^2/\nu$ so that equation \ref{eq:ptuskin_D} predicts a diffusion coefficient that is independent of velocity.

In addition to the result in equation \ref{eq:ptuskin_D} valid in the
limit  $\tau_{\mathrm{diff}} \ll \tau_w$, Ptuskin also
derived the diffusion coefficient for when $\tau_w 
\ll \tau_{\mathrm{diff}}$  i.e., for $v^2 \ll v_w l \nu$ (still
assuming $\lambda_{\rm mfp} \ll l$).
In this regime,  the magnitude of $\delta y$ generated by the waves is  
$\delta y \sim \left(\diver{u}\right) \tau_w/3 \sim u_{l,c} / v_w$, where $\tau_w
\sim l/v_w$ is the wave period and $u_{l,c}$ is the amplitude of
compressible component of the velocity on a given scale $l$.
The fluctuations in momentum with amplitude $\delta y$ are reversible until particles diffuse out of a given wave into an uncorrelated fluctuation, which happens  on the timescale $\delta t \sim \tau_{\mathrm{diff}} \sim l^2 / \kappa$.
The resulting diffusion coefficient is thus 
\begin{equation}
  \label{eq:ptuskin_D_B}
  D \simeq v^2 \frac{2 u_{l,c}^2 \kappa}{9 v_w^2 l^2}  \ \ \  (\tau_{\rm diff} \gg \tau_w).
\end{equation}
This is consistent with the third inequality in eq. 28 of \citet{Ptuskin1988a} up to numerical factors of order unity.   For a scattering rate $\nu$ that is independent of velocity, equation  \ref{eq:ptuskin_D_B} implies that eddies of a given scale $l$ contribute a velocity diffusion coefficient $D \propto v^4$.   The net velocity diffusion coefficient in fully developed turbulence depends on the relative contribution of eddies on different scales.     For density fluctuations associated with slow magnetosonic waves at $\beta \lesssim 1$, which roughly follow an anisotropic critical balanced cascade \citep{Goldreich1995}, this leads to a velocity diffusion coefficient $D \propto v^2$ \citep{Chandran2003a}. 

The results summarized here for the velocity diffusion coefficients produced by rapid pitch-angle scattering are consistent with those derived by \citet{Jokipii2010} in the case of rapid spatial diffusion  (our eq. \ref{eq:ptuskin_D}; their eq. 20).   We do not, however, agree with their results for  slow spatial diffusion  (our eq. \ref{eq:ptuskin_D_B}; their eq. 19).  Instead, our results in the limit of slow spatial diffusion are consistent with those originally derived by \citet{Ptuskin1988a}.

For the magnetized plasmas of interest  in this paper, the
spatial diffusion is not isotropic but instead primarily along magnetic field lines. \citet{Chandran2004}
show that the resulting velocity diffusion coefficient differs
somewhat from that of \citet{Ptuskin1988a}; we briefly reproduce
their argument here, focusing for simplicity on the case of rapid diffusion with $\tau_{\rm diff} \ll \tau_{\rm corr}$.   Because of the strong large-scale magnetic field, particles undergo a one-dimensional random walk along field lines and
are likely to return to their original eddy several
or more times.  As a result, the correlation time for the particle-turbulence
interaction is  the decorrelation time of the turbulence,
$\tau_{\mathrm{corr}}$, rather than the diffusion time across an eddy.  In one correlation time, a typical particle will
traverse $N \sim \left(\kappa_\parallel
  \tau_{\mathrm{corr}}\right)^{1/2} / l_D$ large-scale eddies, where
$\kappa_\parallel$ is  the spatial diffusion coefficient along
magnetic field lines. The average velocity divergence felt by the
particle during this longer correlation time is  reduced by a
factor of $N^{1/2}$ because the individual eddies are not correlated
with one another. Thus the rms change in $y$ over a correlation time
is given by 
\begin{equation}
  \label{eq:chandran_dy}
  \begin{split}
    \delta y & \sim \frac{1}{3} \int_o^{\tau_{\mathrm{corr}}} dt ~ \nabla
    \cdot \mathbf{u} \\ & \sim \frac{1}{3} \tau_{\mathrm{corr}} \left(\nabla
      \cdot \mathbf{u}\right)_{\mathrm{D}} N^{-1/2},
  \end{split}
\end{equation}
and the resulting velocity diffusion coefficient is
\begin{equation}
  \label{eq:chandran_D}
    D_{NR} \sim \frac{v^2 l_D}{18} \left( \nabla \cdot \mathbf{u} \right)_{\mathrm{D}}^2 \sqrt{\frac{\tau_{\mathrm{corr}}}{\kappa_\parallel}},
\end{equation}
consistent with eq. 15 of \citet{Chandran2004}. For a scattering rate $\nu$ that is independent of velocity, equation \ref{eq:chandran_D} predicts $D \propto v$, in contrast to the $D \propto v^0$ prediction of equation \ref{eq:ptuskin_D} for isotropic diffusion.

Equation \ref{eq:chandran_D} relies on the
fact that for high gyrofrequency particles, spatial diffusion is primarily along field lines,
and thus the particles  undergo a 1D random walk, rather than a
fully 3D random walk. This anisotropy, however, does not significantly
affect the diffusion coefficient in the $\tau_{{\rm diff}} \gg \tau_w$ limit,  given in equation \ref{eq:ptuskin_D_B}, except that the eddy size must be interpreted as a parallel correlation length. 
For $\beta \gg 1$, even this is of little practical importance given that the compressible (fast mode)
component of MHD turbulence cascades fairly isotropically \citep{Kowal2010a}.

\subsubsection{Non-Resonant Diffusion:   Moving Scatterers}
\label{sec:moving}

The \citet{Ptuskin1988a} and \citet{Chandran2004} results summarized in the previous section apply if the scattering is isotropic in the local frame of the fluid.   If, instead, the scatterers are randomly moving forward or backward in the fluid frame with a velocity $v_s$ (e.g., high-frequency Alfv\'{e}n waves), and the particles scatter elastically in the scattering frame,  non-relativistic particles will experience
a change in velocity of  $\delta v \sim v_s$ due to the scattering.   Given a scattering rate 
 $\nu$, the scattering will produce an additional contribution to the velocity diffusion of 
 \begin{equation}
 \label{eq:moving_scat}
D_{\rm Fermi} \sim v_s^2 \nu \ \ \ \ ({\rm moving \ scatterers}).
\end{equation}
This is essentially
Fermi scattering \citep{Fermi1949a}  in the non-relativistic
limit.     Note that because the velocity diffusion coefficient produced by interaction with turbulent density compressions scales as $D \propto v$ at high velocities and high scattering rates (eq. \ref{eq:chandran_D}), the latter will dominate over the Fermi-like diffusion coefficient estimated in equation \ref{eq:moving_scat} at sufficiently high velocities.

An important distinction between scatterers that are stationary versus
those moving relative to the fluid frame lies in the source of energy
for the resulting velocity diffusion.  Scattering that is elastic in
the fluid frame does not transfer any energy to the particles and so
the ultimate energy source for the velocity diffusion is the
turbulence itself.  By contrast, scattering that is elastic in a frame
moving relative to the fluid frame leads to energy transfer from the
scatterers to the test particles independent of the existence of
turbulence.  In our calculations with moving scatterers we specify a
constant scattering rate $\nu$ and there is no back reaction on the
scatterers associated with this energy transfer.

\section{Numerical methods}
\label{sec:numericalMethods}

Our simulations consist of charged test particles evolving in
the macroscopic electric and magnetic fields of isothermal, subsonic
MHD turbulence. Apart from our addition of explicit pitch-angle
scattering, which we describe below, our computational approach is
identical to that of \cite{Lynn2012}. Dimensional quantities
throughout the paper are expressed in units of the sound speed $c_s$ and the driving scale
$L$, when not explicitly stated (because we drive at multiple $k$, the actual driving scale is slightly smaller than $L$; see Table \ref{table:fiducial}).

\subsection{Turbulence simulations}
\label{sec:turbSims}

We simulate ideal MHD turbulence with the Athena code
\citep{Stone2008}. The velocity field of the turbulence is driven
solenoidally using an Ornstein-Uhlenbeck process, 
which has a characteristic autocorrelation time
$t_{\rm OU}$. Fiducial
properties for the MHD simulations used in this work are summarized in Table 
\ref{table:fiducial}, and any simulation with different parameters is
explicitly noted.   Our fiducial simulation parameters are broadly similar to those measured in the near-Earth solar wind.

In our calculations, we choose a correlation time similar to the outer  
scale eddy turnover time, to mimic a natural driving process. The
simulation box is extended in the parallel 
direction along the mean magnetic field, because otherwise the particles (which obey periodic
boundary conditions) would unphysically interact with the same eddies multiple times before
the eddies decorrelate.

\begin{table}
  \begin{center}
    \caption{Summary of fiducial simulation properties}
    \begin{tabular}{ c c }
      \\
      \hline \hline
      Parameter & Value \\
      \hline \hline
      Resolution & $1024\times128^2$ \\
      Volume ($L^3$) & $16 \times 2^2$ \\
      $\dot{\epsilon}$ ($c_s^3 / L$)\footnote{The turbulent energy input
        rate. This corresponds to a sonic Mach number of $\simeq 0.35$.} & 0.1 \\
      $\beta$\footnote{Ratio of thermal to magnetic pressure.} & 1 \\
      $t_{\rm OU} \, (L/c_s)$  & $1.5$ \\
      $\tau_{\mathrm{corr}} \, (L/c_s)$\footnote{$t_{\rm OU}$
        refers to the correlation time in the Ornstein-Uhlenbeck turbulence forcing, while
        $\tau_{\mathrm{corr}}$ is the measured correlation
        time in the saturated state of the turbulence. We choose $t_{\rm OU} > \tau_{\mathrm{corr}}$ so that the driving does not artificially reduce the 
        correlation time.} & $0.14$\\
      $\left( \diver{u} \right)_{\mathrm{D}} \, (c_s/L)$ & $0.88$\\
      $l_D$ ($L$)\footnote{Outer (driving) scale of the turbulence.} & $0.39$\\
      $\delta n/n$\footnote{RMS density fluctuations in the saturated
        turbulence.} & $0.23$ \\
      $\delta B/B_0$\footnote{RMS magnetic field fluctuations in the
        saturated turbulence.} & $0.28$ \\ 
            $N_{\mathrm{particles}}$ & $2^{11} \times 10^3 \simeq 2 \times 10^6$ \\
      $\Omega_0$ ($c_s/L$)\footnote{Test particle gyrofrequency.} & $2
      \times 10^5$ \\
      
         \end{tabular}
    \label{table:fiducial}
  \end{center}
\end{table}

\subsubsection{Measurement of turbulence properties}
\label{sec:turbulenceProperties}
The predicted efficiency of non-resonant velocity diffusion, summarized
in equation 
\ref{eq:chandran_D}, depends on two properties of the turbulence,
$\left( \diver{u}\right)_{\mathrm{D}}$ and
$\tau_{\mathrm{corr}}$. To order of magnitude, these can be approximated as
$u_{\mathrm{rms}}/l_D$ and $l_D/u_{\mathrm{rms}}$, respectively, for our $\beta \sim 1$ turbulence. However, it is more accurate to measure these two parameters directly
in the turbulence.

Recall that $\left( \diver{u} \right)_{\mathrm{D}}$ refers to the velocity
divergence of the largest eddies. To measure this, we
first Fourier transform the velocity field and then multiply by a
lowpass window function of the form $\exp{\left[-C \left( k l_D / 2 
      \pi\right)^2\right]}$, where $C=\ln{2}$ is chosen so that the
window function is $1/2$ at $k/2\pi = l_D$. $l_D$ is defined by $l_D \equiv
\langle 2 \pi/k \rangle$, where the average is taken over the power
spectrum of the driving. Our driving power spectrum is a power-law in
$k$ with a 1D power spectrum of $k^{-3}$ defined between
$k_{\mathrm{min}} = 2\pi/(L/2)$ and $k_{\mathrm{max}} = 2\pi/(L/4)$, which
gives $l_D \simeq 0.39 L$. We then calculate the rms velocity divergence,
which may be done easily in Fourier- or real-space.   This yields
$\left( \diver{u} \right)_{\mathrm{D}} \simeq 0.88 c_s/L$ for our
fiducial simulation summarized in Table \ref{table:fiducial}.

To calculate the correlation time, we choose one point per
cpu to output a time-series of the velocity magnitude. The point
chosen is fixed with respect to the origin of the local grid. Because
the local grids periodically tile the entire domain, these points are
arranged in a lattice. There are 512 points in our fiducial case,
which should be approximately independent, as there is approximately
one per physical eddy volume. For 
each of these lattice points, we calculate the autocorrelation of the
time series in using the |sfsmisc| package in R
\citep{sfsmisc,RLang}. These autocorrelation functions are
then averaged over all lattice points, and fit with a functional form 
$\exp{\left(-t/\tau_{\mathrm{corr}}\right)}$, which gives
$\tau_{\mathrm{corr}} \simeq 0.14 L/c_s$ for our fiducial 
simulation (see Table \ref{table:fiducial}). Quantitatively, this is
roughly consistent with an estimate of $(k_{drive} \, \delta
v_{\mathrm{rms}})^{-1} \simeq 0.18 \, L/c_s$.

Finally, we can decompose the turbulence into components corresponding
to the linear MHD Alfv\'{e}n, slow, and fast modes, following the
Fourier space method of \citet{Cho2003}. This method makes the
simplifying assumption that the magnetic field points along an axis of
the simulation domain, which
will be asymptotically appropriate as $\delta B/B_0 \rightarrow 0$.  For our fiducial simulation summarized in Table \ref{table:fiducial}, we
find that 50\%, 45\%, and 5\% of the turbulent energy is in the
Alfv\'{e}n, slow, and fast modes respectively.   This is somewhat more
slow  and fast mode energy than measured in the near-Earth solar wind
\citep{Howes2012}, likely because of the strong plasma heating (and
thus turbulent dissipation) produced by the compressible component of
the turbulence (see \S \ref{sec:particleTransport}).   An alternative
decomposition of the turbulence is into solenoidal and compressive
($\propto \nabla \cdot v$) fluctuations:  we find that $\sim 10 \%$ of
the energy is in compressive fluctuations.   This contains both slow
and fast mode contributions for our $\beta = 1$ turbulence.    The
corresponding rms density fluctuation is $\delta n/n \simeq 0.23$,
somewhat larger than the {\em in situ} value of $\delta n/n \sim 0.1$
measured in the solar wind (e.g., \citealt{Tu1995}).

\subsubsection{Effect of limited dynamic range}
\label{sec:dynamic_range}

\begin{figure}
 \includegraphics[width=\figwidth]{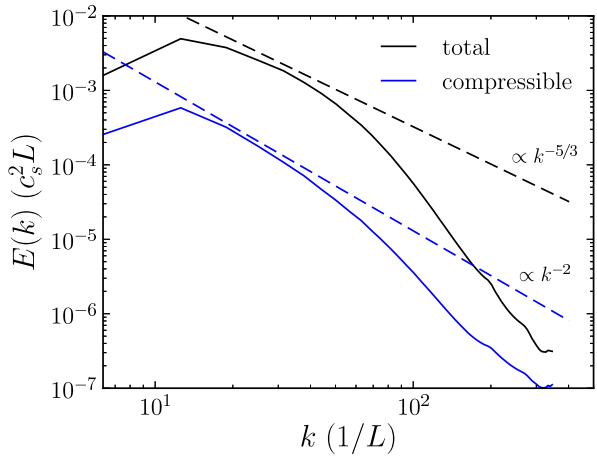}
  \caption{1D angle-averaged kinetic energy power spectra in our
    turbulence 
    simulations. Simulating many ``outer-scale'' eddies in the
    parallel direction limits the inertial range of our turbulence to
    less than a decade.}
  \label{fig:spectrum}
\end{figure}

Given our periodic boundary conditions and a box of finite size, a
test particle will cross the box multiple times.  If a particle
returns to nearly the same location following a magnetic field line,
before the turbulence has had time to randomize, it will experience
artificial correlations in the turbulence and therefore artificial
acceleration along its trajectory. To minimize this effect, we extend
the box in the parallel direction, so that particles interact with
many uncorrelated eddies as they cross the box. However, for fixed
computational time, this effectively means that we are reducing the
dynamic range of the simulation in terms of the number of simulation
elements that comprise each turbulent eddy.   Figure
\ref{fig:spectrum} shows the one-dimensional kinetic energy power
spectrum of the turbulence in our fiducial simulation.  A 
clear consequence of extending the box along the mean magnetic field
is that the inertial range of the turbulence is modest, less than a
decade in k.   We have found, however, that not extending box along
the mean field significantly changes the acceleration of high energy
particles, so this compromise of limited dynamic range is necessary.

Our analytical estimates in \S \ref{sec:transportProperties} are given
in terms of arbitrary power spectra for the turbulent fluctuations. In
many cases, the diffusion and acceleration of particles in the
presence of pitch-angle scattering is dominated by interactions with a
particular scale in the turbulence, typically either the smallest or
largest scale in the inertial 
range. In this case, simulating a large inertial range may not be
critical, though it would obviously be preferable. Later in the paper
we highlight the specific results that we suspect may be most
influenced by our limited dynamic range.

\subsection{Test particle integration}
\label{sec:testParticles}
Our particle integration methods are described in detail in 
\citet{Lehe2009}. Particles are initialized in the fully saturated
turbulence, and then evolved according to the Lorentz force. The $E$
and $B$-fields are those on the MHD grid, interpolated to the
particle's location using the triangular-shaped cloud
\citep{Hockney1981} method in space and time. Particles are integrated
using the \citet{Boris1970} particle pusher, which is sympletic and
symmetric in time, and conserves energy and the magnetic moment
adiabatic invariant to machine precision in tests with constant
fields.    We  initialize particles randomly across the simulation
domain with the initial values of $v_\perp$ and $v_\parallel$ defined
relative to the local rest-frame of the MHD fluid in the
simulation. More details are provided in \citet{Lehe2009}. 

\subsubsection{Pitch-angle scattering}
\label{sec:scatterMethod}
We implement pitch-angle scattering by specifying a scattering rate,
$\nu$.    We consider two cases to bracket the likely range of physical scattering mechanisms.   The first is when the scatterers are at rest in the fluid frame.   This seems likely to be appropriate for scattering by mirror or firehose instabilities, but not by, e.g., Alfv\'en waves, which move relative to the fluid frame at the Alfv\'en speed.   We consider  scatterers moving along the local magnetic field direction to model the latter.

At each scattering timestep $dt_s$ (defined below), and for each test particle, we subtract off
 the local fluid velocity, to perform the scattering in the
fluid frame. For the simulations with moving scatterers, we move into
the frame of the scatterer by subtracting off a further component of
the velocity $\pm v_s \hat{b}$, where $\hat{b}$ is the local magnetic
field direction and the scatterer is randomly chosen to be moving
forward or backward in the fluid frame. We choose an isotropic random
unit vector $\hat{e}_1$. We then 
calculate $\hat{e}_2 = \hat{e}_1 \times \hat{v}$, i.e., a unit vector
perpendicular to the particle velocity $\mathbf{v}$. We rotate the
test particle's velocity vector around $\hat{e}_2$ through a random 
angle uniformly distributed between 0 and
$\theta_{\mathrm{max}}$. $\hat{e}_2$ is chosen to be perpendicular to
$\mathbf{v}$ so that the rotation of $\hat{v}$ occurs along a great
circle on the unit sphere. We
choose the timestep for the scattering 
procedure according to $dt_s = \theta_{\mathrm{max}}^2/(6 \nu)$. We
implement this rotation using 
Rodrigues' rotation formula. Finally, the local fluid
velocity
(in addition to any scatterer velocity as appropriate)
is added back to the particle's velocity to return us to the lab frame.

This procedure ensures that the pitch-angle scattering rate is in fact given by our input parameter
$\nu$ as long as $\theta_{\mathrm{max}} \ll 2 \pi$, so that scattering
is dominated by small changes in pitch-angle. We typically choose
$\theta_{\mathrm{max}} = 0.6$, which is sufficiently small that
diffusion coefficients differ by less than 1\% from their
saturated values at $\theta_{\mathrm{max}} \rightarrow 0$, while
minimizing computational
effort associated with frequent scatterings. 

In addition to diffusion in velocity space, pitch-angle
scattering  also
leads to spatial diffusion with a mean free path
$\lambda_\parallel \sim v / \nu$ and a diffusion coefficient
$\kappa_\parallel = (2/3)  v^2 / \nu$, where the parallel subscript
indicates that the spatial diffusion is still primarily along magnetic
field lines in our calculations.  In particular, perpendicular spatial
  diffusion $\kappa_\perp$ is  $\sim r_G^2 \nu$, where $r_G$ is the
particle gyroradius. We simulate
  high-gyrofrequency particles with 
small $r_G$ and so perpendicular
diffusion is negligible even at the highest $\nu$ we consider.

\subsection{Calculating Velocity Diffusion Coefficients}

\begin{figure}
 \includegraphics[width=\figwidth]{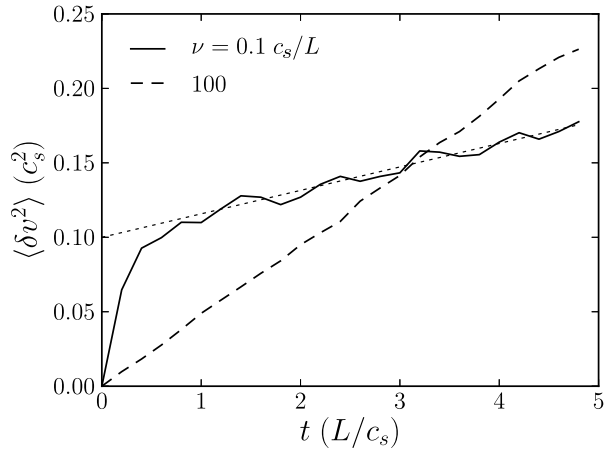}
  \caption{Dispersion in velocity, measured relative to the fluid frame, as a function of time for test particles
     that were initially delta-functions with $v_i \simeq
    0.025\,c_s$.  The turbulence has properties given by Table \ref{table:fiducial}.  For the low-scattering run there is an initial ``jump'' in velocity dispersion caused by the changing 
    local parallel fluid velocity relative to the particles, after
    which $\langle \delta v^2 \rangle (t)$ increases linearly,
    consistent with diffusion. The high-scattering run has no such
    jump because the particles remain tied to the local fluid
    frame by the rapid scattering. The dotted 
    curve shows the fit we use to 
    calculate diffusion coefficients in the case with the initial jump.}
  \label{fig:deltaVs}
\end{figure}

The  velocity diffusion
coefficients are calculated according to the formal definition
\begin{equation}
  \label{eq:diffusionDefinition}
  D \equiv \frac{\langle \delta v^2 \rangle} {2 \, \delta t},
\end{equation}
where the average is over many particles with the same initial
velocity. However, one subtlety is that a particle which experiences no acceleration can
nevertheless appear to change velocity when measured with respect to
the local fluid frame.   This is because the fluid's parallel velocity can
change even if a given test particle feels no electromagnetic acceleration.   Thus, absent pitch-angle scattering, after a time of
order the correlation time of the fluid, an ensemble of particles initially at rest relative to the fluid
 will develop a finite dispersion in parallel velocity given by $\delta v \sim
v_{\parallel,\mathrm{rms}}$ of the turbulence. (This effect becomes less significant at high pitch-angle scattering rates because the particles are `pinned' near the location in the fluid where they are initialized.)   Figure \ref{fig:deltaVs} plots two representative $\langle \delta v^2
\rangle (t)$ curves and shows a rapid increase in $\langle \delta v^2 \rangle (t)$ at early times for the low scattering rate calculation.   To not include this  in our diffusion coefficient calculations, we measure 
diffusion coefficients by instead fitting a straight line after the
initial jump using least-squares (as indicated in Fig. \ref{fig:deltaVs}).

\section{Test-particle Diffusion and Heating}
\label{sec:particleTransport}

\subsection{Velocity Diffusion:  Stationary Scatterers}
\label{sec:tpResonantDiffusion}

\begin{figure}
  \includegraphics[width=\figwidth]{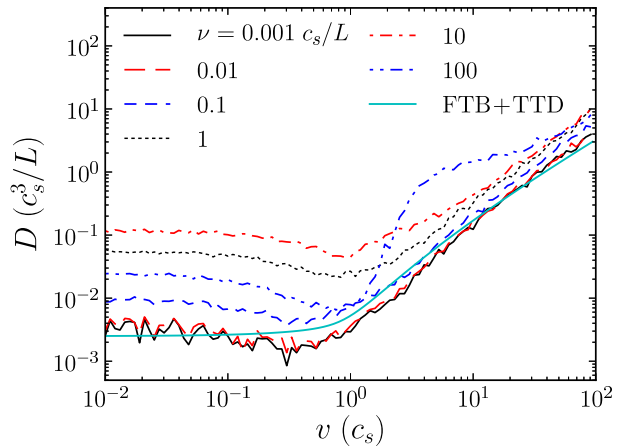}
  \caption{Test particle velocity diffusion 
    coefficients for low scattering rates for the fiducial simulation (see Table \ref{table:fiducial}).  Scatterers are stationary in the fluid frame.  The
    solid cyan curve is a fit to a functional form given by
    a sum of Fermi Type-B and resonance-broadened transit-time
    damping in which particles interact with turbulent fluctuations in magnetic field strength. This provides a reasonable description at low scattering rates and/or high velocities.   At high scattering rates ($\nu \sim 100 c_s/L$), however, the functional form of the diffusion coefficient in the test particle calculations changes significantly.  This corresponds to the transition to a non-resonant regime in which particles interact primarily with turbulent density fluctuation (see Fig. \ref{fig:D_highNu} and \S \ref{sec:ptuskinA}). }  
  \label{fig:D_lowNu}
\end{figure}

\begin{figure}
  \includegraphics[width=\figwidth]{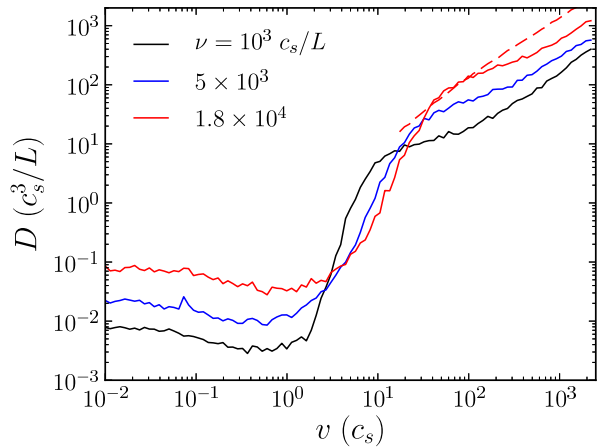}
  \caption{Test particle velocity diffusion
    coefficients for the fiducial simulation at high scattering rates (solid curves), corresponding to the non-resonant regime discussed in \S \ref{sec:ptuskinA}.   Scatterers are stationary in the fluid frame.   For high
    velocities, the numerical results are reasonably well-approximated by eq. 
    \ref{eq:chandran_D} (dashed curve), which describes velocity diffusion by non-resonant interactions with
    turbulent density fluctuations.   The rapid increase in the diffusion coefficient with velocity at intermediate velocities $\sim {\rm few} \, c_s$ can produce a non-thermal tail in the distribution function (see Fig. \ref{fig:powerLawTails}).
    } 
  \label{fig:D_highNu}
\end{figure}

Figures \ref{fig:D_lowNu} and
\ref{fig:D_highNu} show diffusion coefficients for a broad range of pitch-angle
scattering rates $\nu$ for our fiducial simulation (Table \ref{table:fiducial}); the scattering  is isotropic in the fluid frame.    The chosen values of $\nu$ include  scattering times significantly longer than, and significantly shorter than, the outer-scale turnover time of the turbulence.  At the lowest scattering rates ($\nu \sim 10^{-3}-10^{-1} \, c_s/L$, corresponding to scattering times somewhat smaller than the nonlinear timescale of the turbulent fluctuations),
the diffusion coefficient is well approximated by the pitch-angle average of the parallel diffusion coefficient appropriate for resonance broadened TTD + Fermi Type-B diffusion described  in \S \ref{sec:resonantDiffusion} (cyan line in Fig. \ref{fig:D_lowNu}).  In this limit, scattering isotropizes the velocity-space diffusion but does not fundamentally alter the physics of how particles interact with the turbulent fluctuations.

For scattering rates  $0.1 \, c_s/L \lesssim \nu \lesssim 10 \, c_s/L$, the diffusion coefficient increases at low velocity but not significantly for $v \gtrsim {\rm few} \, c_s$.   The reason is that large-scale eddies produce the Fermi Type-B diffusion at low velocities -- these interactions are thus fundamentally altered once the scattering rate is comparable to the frequency of the large-scale eddies.  By contrast, eddies of all scales (including, in particular, small scales) contribute significantly to the high velocity TTD diffusion.   The small scale eddies have higher nonlinear frequencies and thus a higher scattering rate is needed to modify the velocity-space diffusion of high speed particles.

\begin{figure}
  \includegraphics[width=\figwidth]{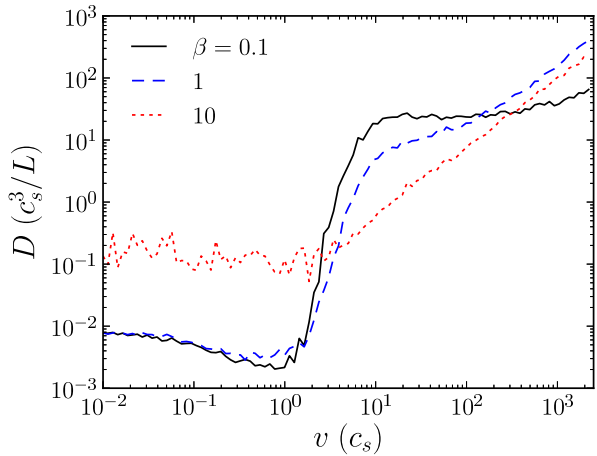}
  \caption{Test particle velocity diffusion
    coefficients for $\nu = 10^3 c_s/L$ for three different $\beta$, corresponding to the non-resonant regime discussed in \S \ref{sec:ptuskinA}.   Scatterers are stationary in the fluid frame.   The results are similar for low $\beta \lesssim 1$ where slow modes dominate the compressional fluctuations.   For  $\beta = 10$ the diffusion coefficient resembles the $\beta = 1$ results at much lower scattering rates (see Fig. \ref{fig:D_lowNu}).  This is because the compressional energy is significantly lower in the $\beta = 10$ simulation, which decreases the efficiency of the non-resonant diffusion.} 
  \label{fig:D_highNubeta}
\end{figure}

For $\nu \gtrsim 100 \, c_s/L$, there is a qualitative change in the
properties of the velocity diffusion coefficients in Figures
\ref{fig:D_lowNu} and 
\ref{fig:D_highNu}, particularly for particles with $v \gtrsim c_s$.
This corresponds to the onset of non-resonant diffusion governed by
interactions between particles and turbulent density fluctuations.
The properties of the high scattering rate diffusion depend on the
ratio of the diffusion time to the wave period
$\tau_{\mathrm{diff}}/\tau_w \propto \nu/v^2$.   Lower velocity
particles are in the regime where $\tau_{\rm diff} \gtrsim \tau_w$ and
equation \ref{eq:ptuskin_D_B} predicts that the velocity diffusion
coefficient is a strong function of velocity, with $D \propto v^4$.
This is consistent with the numerical results in  Figures
\ref{fig:D_lowNu} and \ref{fig:D_highNu} for high scattering rates and
intermediate velocities $v \sim 2-10 \, c_s$.    Note that we do not
numerically recover \citet{Chandran2003a}'s prediction that
compressible slow mode turbulence at $\beta \lesssim 1$ should produce
a $D \propto v^2$ scaling due to the contribution of eddies of
different scale $l$ in equation \ref{eq:ptuskin_D_B}. This may be
because the finite dynamic range in our simulations means that we do
not capture the full inertial range of the turbulence (this is
exacerbated by our anisotropic box [see Table \ref{table:fiducial}],
which is required in order to avoid artifacts due to periodic boundary
conditions along the mean field).  In the limit that eddies of a
single spatial scale dominate the velocity diffusion, equation
\ref{eq:ptuskin_D_B} implies $D \propto v^4$, consistent with our
numerical results.

For yet higher velocity particles, $\tau_{\rm diff} \lesssim \tau_w$ and equation \ref{eq:chandran_D} predicts $D \propto v$ in the rapid scattering limit.   This is consistent with the numerical results in Figures \ref{fig:D_lowNu} and \ref{fig:D_highNu}.    Quantitatively, 
 the non-resonant regime of
equation \ref{eq:chandran_D} 
requires simultaneously satisfying $l_D \gg \lambda_{\mathrm{mfp}}=
v/\nu$ and 
$\tau_{\mathrm{diff}} \ll \tau_w$, which restricts us to the
following range of velocities:
\begin{equation}
  \label{eq:vel_range}
  \nu l_D^2 / \tau_{\mathrm{w}} \ll v^2 \ll \nu^2 l_D^2.
\end{equation}
For the simulation with $\nu = 5 \times 10^3 \, c_s/L$, the
high-velocity non-resonant diffusion is appropriate in our test particle calculations for $73 \ll v/c_s
\ll 1950$, while for $\nu = 1.8 \times 
10^4 \, c_s/L$, particles with $140 \ll v/c_s \ll 5700$ should satisfy equation \ref{eq:chandran_D}. In Figure \ref{fig:D_highNu}, we include the
prediction of equation \ref{eq:chandran_D} for the $\nu = 1.8 \times 10^4 \, c_s/L$ simulation as a dashed curve (measuring $\nabla \cdot u$ and $\tau_{\rm corr}$ as  in \S \ref{sec:turbulenceProperties} to  determine the predicted normalization of $D$). The
velocity diffusion is reasonably well approximated by equation
\ref{eq:chandran_D} in the appropriate velocity interval.

Figure \ref{fig:D_highNubeta} shows how the diffusion coefficient in the high scattering rate, non-resonant regime depends on $\beta$ (for fixed $\nu = 10^3 c_s/L$).   The turbulence has the same driving as in the fiducial simulation in Table \ref{table:fiducial}.   The key difference in the resulting turbulence properties  is that the fraction of the turbulent energy in compressional fluctuations (those $\propto \nabla \cdot v$; see \S \ref{sec:turbulenceProperties}) is only $\sim 1.5 \%$ for $\beta = 10$ versus $10 \%$ and $17 \%$ for $\beta = 1$ and $0.1$, respectively.  This is because fast modes dominate the compressional fluctuations at high $\beta$ and are not efficiently excited in subsonic  turbulence with solenoidal driving.

The results for the diffusion coefficient in Figure \ref{fig:D_highNubeta} are  similar for $\beta = 1$ and $\beta = 0.1$.   This is consistent with the fact that at low $\beta$ slow magnetosonic modes dominate the compressional fluctuations and are energetically important in  MHD turbulence.   By contrast, the diffusion coefficient changes significantly at $\beta \sim 10$, where fast modes dominate the compressional fluctuations.  In particular, the $\beta = 10$ results with $\nu = 10^3 \, c_s/L$ in Figure \ref{fig:D_highNubeta} are similar to the $\beta = 1$ results at a much lower $\nu \sim 10 \, c_s/L$ (see Fig. \ref{fig:D_lowNu}).   This is because of the much lower compressional energy in the $\beta = 10$ simulation, which decreases the efficiency of the non-resonant diffusion  (see \S \ref{sec:ptuskinA}).

\subsection{Velocity Diffusion:  Moving Scatterers}
\label{sec:moving_scatterers}

\begin{figure}
\includegraphics[width=\figwidth]{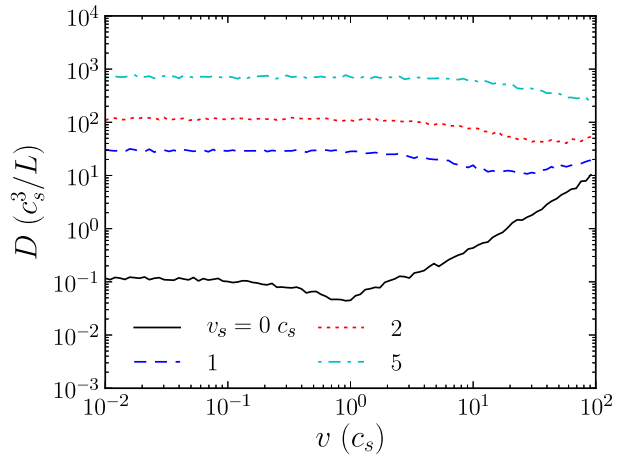}
  \caption{Test particle velocity diffusion 
    coefficients for the fiducial simulation with $\nu=10 \, c_s/L$ and scatterers
    moving at various velocities $v_s$. The moving scatterers cause
    Fermi-like velocity diffusion consistent with $D \sim v_s^2 \nu$
    (see eq. \ref{eq:moving_scat} and \S \ref{sec:moving}).}  
  \label{fig:D_moving}
\end{figure}

Figure \ref{fig:D_moving} shows test particle results from
simulations with moving scatterers. The velocity diffusion coefficient is nearly independent of particle velocity.   These results are consistent with Fermi-like scattering predicted by equation \ref{eq:moving_scat}, namely  $D \sim v_s^2 \nu$.   At sufficiently high velocity, the non-resonant diffusion $D \propto v$ shown in Figures \ref{fig:D_lowNu} and \ref{fig:D_highNu} would eventually dominate over the roughly constant diffusion coefficient shown in Figure \ref{fig:D_moving}.

\subsection{Test-particle heating}

\begin{figure}
  \includegraphics[width=\figwidth]{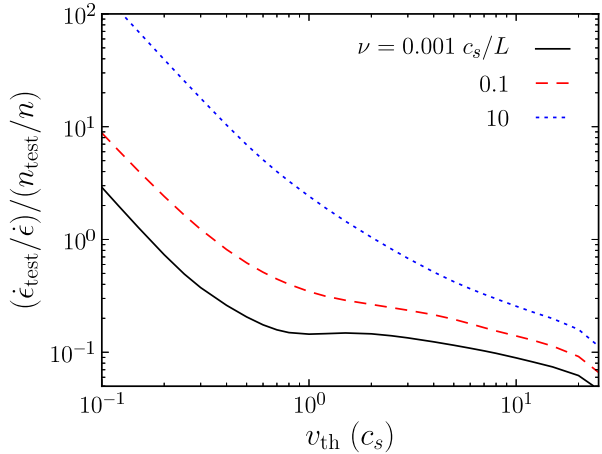}
  \caption{Particle heating rate $\dot \epsilon_{\rm test}$, normalized by the turbulent cascade rate
  $\dot{\epsilon}$, as a function of thermal velocity $v_{\rm th}$, for an
  initially Maxwellian distribution function.   The heating rate is measured from $t=2.5-20
  L/c_s$ in the fiducial simulation and is $\propto n_{\rm test}/n$, where $n_{\rm test}/n$ is the number density of test particles relative to the fluid density in the simulation.    
  Modest pitch-angle scattering rates $\nu$ enforce
  isotropy and significantly enhance the turbulent heating of the plasma.} 
  \label{fig:nusHeating}
\end{figure}

Figure \ref{fig:nusHeating} shows how the pitch-angle scattering
rate $\nu$ affects the heating of test particles that have an initially
Maxwellian distribution with thermal speed $v_{\rm th}$.   The $\nu = 10^{-3} c_s/L$ results shown are nearly identical to the heating rates in the absence of pitch-angle scattering.  
For the calculations in Figure \ref{fig:nusHeating}, we assume that the scattering is at rest in the local fluid frame.   Note that for our fiducial simulation parameters (Table \ref{table:fiducial}), protons  correspond to $v_{\rm th}  = c_s/\sqrt{2}$, electrons to $v_{\rm th} \simeq 30 c_s$ and minor ions (with temperatures of order the proton temperature) to $v_{\rm th} \lesssim c_s$ in Figure  \ref{fig:nusHeating}.   

Not surprisingly, the enhanced velocity diffusion shown in Figures \ref{fig:D_lowNu} and \ref{fig:D_highNu} leads to a corresponding increase in the net particle heating rate.   At the scattering rates we focus on here, this increase in the heating rate is a consequence of pitch-angle scattering enforcing isotropy, which leads to a substantially enhanced TTD heating rate at $v \gtrsim c_s$ and an enhanced FTB heating rate at low velocities.    Quantitatively, Figure \ref{fig:nusHeating} shows that even a modest scattering rate, with $\nu^{-1}$ comparable to the outer-scale turbulent correlation time ($\sim 0.1 L/c_s$; see Table \ref{table:fiducial}), can increase the net heating rate significantly so 
that a large fraction of the compressible turbulent energy can go 
into heating protons.   Moreover, when $\dot \epsilon_{\rm test} \gtrsim \dot \epsilon$ for protons,  our test particle assumption breaks down because the protons will absorb the majority of the compressible turbulent power on large scales in the turbulent cascade.   Figure \ref{fig:nusHeating} also shows that minor ion heating is particularly effective if the ions and protons have comparable temperatures, so that the minor ions have lower thermal velocities.   This is true, e.g., in the slow speed solar wind but not in the fast solar wind, where the ion and proton thermal velocities are comparable \citep{Bochsler2007}; in this case, our results imply comparable proton and minor ion heating rates per particle.

\section{Power-law tails?}
\label{sec:powerLaw}

\begin{figure}
  \includegraphics[width=\figwidth]{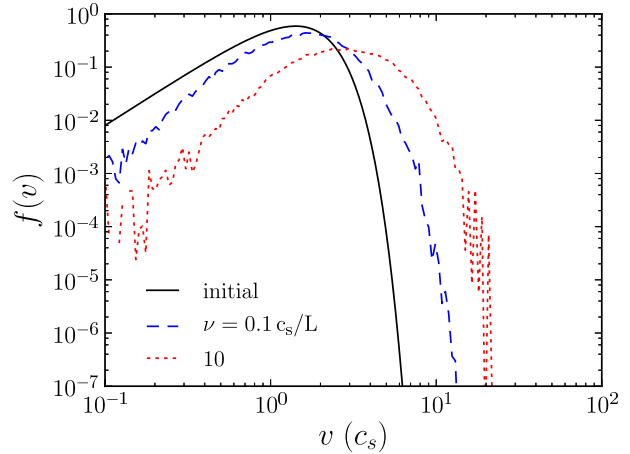}
  \caption{An initially Maxwellian distribution
    of test particles with thermal velocity $v_{\rm th} = c_s$
    is simulated for $t = 20 L/c_s$ (many eddy turnover times) in our fiducial simulation (Table \ref{table:fiducial}),
    with two different values of the imposed scattering rate $\nu$.  The final distribution functions are slightly non-Maxwellian but do not develop a significantly non-thermal tail. $f(v)$
    shown here is the one-dimensional distribution
    function, normalized such that the density is $n_0 = \int dv f(v)$.}
  \label{fig:thermalDist}
\end{figure}

Figure \ref{fig:thermalDist}  shows the late-time distribution function that results when we evolve test particles which initially have a Maxwellian with $v_{\rm th} = c_s$ in our fiducial turbulence simulation (Table \ref{table:fiducial}), for several different scattering rates.   The evolution is for $\sim 20 \, L/c_s$, which is many eddy turnover times.     These results are very similar to the analogous results without pitch-angle scattering in our earlier work \citep{Lynn2012}.   In particular, Figure \ref{fig:thermalDist} demonstrates that at low to moderate scattering rates the particles are efficiently heated by the turbulence (more-so with increasing $\nu$; see Fig. \ref{fig:nusHeating}), but  no significant non-thermal power-law tail develops.   Recall that this  scattering rate regime corresponds to the particles gaining energy from turbulent magnetic compressions  (as in linear TTD of fast and slow magnetosonic waves).   

\begin{figure}
  \includegraphics[width=\figwidth]{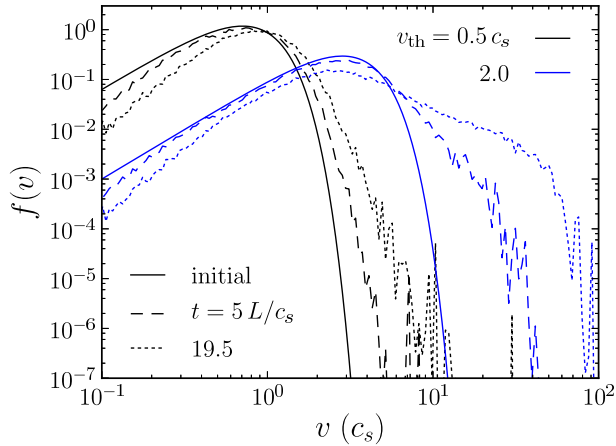}
  \caption{Initially Maxwellian distributions
    of test particles with $v_{\rm th} = 0.5$ and $2 \, c_s$
    are simulated for $t = 20 L/c_s$ (many eddy turnover times),
    with a high scattering rate $\nu = 10^3 \, c_s/L$. We show the
    initial (solid)
    and final (dotted) distributions, as well as at an intermediate
    time (dashed). Power-law tails
    develop in the velocity regime with the enhanced $D \propto v^4$
    velocity diffusion in Fig. \ref{fig:D_highNu}.  There is not, however, a `universal' power-law slope to the resulting distribution function.  Instead, the slope depends on time, thermal velocity of the distribution, and pitch-angle scattering rate (the latter dependence is not explicitly shown here).  $f(v)$
    shown here is the one-dimensional distribution
    function, normalized such that the density is $n_0 = \int dv f(v)$.}
  \label{fig:powerLawTails}
\end{figure}

Figure \ref{fig:powerLawTails} shows  the evolution of the distribution function for $\nu = 10^3 \, c_s/L$ for  two initial thermal velocities, $v_{\rm th} = 0.5$ and $2 \, c_s$ (these two different thermal velocities are included to highlight the physics of the distribution function evolution at high $\nu$; below we also  summarize results for calculations with $v_{\rm th} = c_s$, more relevant for protons).   In the high scattering rate calculations in Figure \ref{fig:powerLawTails},  the resulting distribution function is qualitatively different:   there is a nearly power-law component  over almost a decade in
velocity, particularly for $v_{\rm th} = 2 \, c_s$.   The non-thermal tail also becomes more energetically important at later times, after the distribution function has been evolved for $\sim 20 L/c_s$.   There is no evidence for a universal power-law slope in the velocity distribution function in this regime of rapid pitch-angle scattering.  Instead, the exact slope  depends sensitively on time, the pitch-angle scattering rate, and the thermal velocity of the test particle distribution:  for $v_{\rm th} = 2 \, c_s$, $f \propto v^{-3.5}$ at $t = 5 L/c_s$ and $f \propto v^{-1.5}$ at $t = 20 L/c_s$ while for $v_{\rm th} = 0.5 \, c_s$, $f \propto v^{-7.5}$ at late times and is not convincingly a power-law at earlier times.

The significant non-thermal tail to the distribution function in Figure \ref{fig:powerLawTails} is caused by the steep scaling of the velocity diffusion coefficient with $v$ for intermediate velocities seen in Figure \ref{fig:D_highNu}.\footnote{As noted in \S \ref{sec:tpResonantDiffusion}, \citet{Chandran2003a} argued that for a large turbulent inertial range, compressible slow mode fluctuations at $\beta \lesssim 1$ would lead to a diffusion coefficient $D \propto v^2$, rather than the somewhat steeper result we find numerically (Fig. \ref{fig:D_highNu}).   Even if this were to be the case, we believe that a significant power-law tail would develop, based on analytic and numerical solutions of the Fokker-Planck equation with $D \propto v^2$.}   The `strength' of this power-law tail depends, however, on the assumed pitch-angle scattering rate and the thermal speed of the test particles since this determines how well the $D \propto v^4$ diffusion coefficient regime in Figure \ref{fig:D_highNu} overlaps with the initial Maxwellian of the test particles.  Higher scattering rates lead to $D \propto v^4$ setting in at higher velocity.    This implies significantly less efficient  acceleration of particles out of the thermal population.   Thus a strong power-law tail requires fine-tuning between the scattering rate
and the thermal velocity of the species of interest.
Quantitatively, the power-law tail in Figure \ref{fig:powerLawTails} is much less significant for $v_{\rm th} = 0.5 c_s$ relative to $v_{\rm th} = 2 c_s$.   In addition, in calculations with $v_{\rm th} = 1 \, c_s$ (more appropriate for protons), which are not explicitly shown here,  we find that after $\simeq 5 \, L/c_s$ interacting with our fiducial turbulence (Table \ref{table:fiducial}), the distribution function roughly satisfies $f(v) \propto v^{-3.5}$ for $\nu \simeq 100 \, c_s/L$, $f(v) \propto v^{-6.5}$ for $\nu \simeq 10^3 \, c_s/L$, and $f(v) \propto v^{-10}$ for $\nu \simeq 10^4 \, c_s/L$ over the velocity range $v \sim 3-10 \, c_s$.   This highlights the strong dependence of the resulting distribution function on all of the parameters of the problem.

The absence of a  non-thermal tail at low scattering rates in Figure \ref{fig:thermalDist} is consistent with the general form of the diffusion coefficient in Figure \ref{fig:D_lowNu}, in particular the fact that the diffusion coefficient scales at most $\propto v \log[v]$.    Green's function solutions of the diffusion equation show that only if the diffusion coefficient scales as $D \propto v^2$ or steeper does velocity diffusion alone generically produce a strong non-thermal tail to the distribution function (e.g., \citealt{Gruzinov1999b,Jokipii2010}).   This is  consistent with our numerical results in Figure \ref{fig:thermalDist}.

In his original analysis, Fermi invoked a competition between velocity diffusion and particle escape in order to obtain a power-law  distribution function.  Our simulations do not explicitly have escape and  one might thus worry that the absence of a power-law tail is an artifact of the periodic boundary conditions in our calculations.  This is not the case.    

In the presence of particle escape, the combination of diffusion and escape produces a power-law tail only if the ratio of the acceleration time $\tau_a \equiv v^2/D$ to the escape time $\equiv \tau_e$ is independent of particle energy \citep[e.g.][]{Longair1992}.
In our problem, a well-posed question is whether escape due to spatial diffusion, which is self-consistently produced by pitch-angle scattering, produces the requisite conditions for a power-law tail.   Scattering implies $\kappa \sim v^2/\nu$ so that $\tau_e \propto \nu/v^2$.      If $\nu$ is roughly independent of velocity, only a very steep diffusion coefficient of the form $D \propto v^4$ can produce a power-law tail in the presence of diffusive escape.   This is the regime in which we already find a significant non-thermal tail in Figure \ref{fig:powerLawTails}.   The   velocity diffusion coefficient that results from most of our calculations is (roughly) $D \propto v$, i.e., $\tau_a \propto v$.  This applies in both the quasi-resonant limit and in the high scattering rate limit at high velocities.  Thus, unless for some reason $\nu \propto v^3$, so that $\tau_e \propto v$, weakly compressible MHD turbulence will not generically produce a significant non-thermal tail in the $D \propto v$ regime even in the presence of particle escape.

\section{Conclusions}
\label{sec:conclusions}

We have studied the heating and acceleration of particles by subsonic MHD turbulence by evolving an ensemble of test particles in real time in MHD turbulence simulations.   Our focus in this paper has been on the role of pitch-angle scattering in modifying the efficiency of non-relativistic particle heating and acceleration.   To study this we have implemented explicit pitch-angle scattering (at a fixed rate $\nu$), with the scatterers either at rest in the fluid frame or moving relative to the fluid frame along the local magnetic field. We suspect that these two cases bracket reality in problems of interest, representing the role of mirror/firehose instabilities (which are non-propagating instabilities and thus roughly at rest in the fluid frame) and high frequency ion cyclotron, Alfv\'en, or whistler waves, respectively.   In our calculations the scattering rate is independent of velocity, but this may not be the case in real systems.  In particular, the mechanisms that dominate pitch-angle scattering may differ for thermal and super-thermal particles.   Results for a velocity-dependent scattering rate can be obtained from the calculations presented here by appropriately interpolating between the diffusion coefficients for different scattering rates shown in Figures \ref{fig:D_lowNu}-\ref{fig:D_highNubeta} (this is formally valid for our test particle calculations).

In all of our calculations, we have focused on the evolution of  particles with gyrofrequencies much larger than the resolved frequencies of the turbulent fluctuations.   In the absence of pitch-angle scattering, the magnetic moment of our test particles is thus conserved:   particles only undergo diffusion in $v_\parallel$.    The reason for restricting our analysis to these high gyrofrequency particles is that the outer scale turbulent fluctuations in  astrophysical and heliospheric systems have frequencies much smaller than the particle gyrofrequencies.   Thus the high gyrofrequency limit is the correct limit for studying particle heating and acceleration using MHD simulations.   In addition, much of the energy in MHD turbulence cascades anisotropically in wave-vector space, and there is little power at frequencies comparable to the ion cyclotron frequency, even when the  turbulent fluctuations have wavelengths  comparable to the ion Larmor radii (e.g., \citealt{Schekochihin2009}).  A corollary of this is that it is difficult to predict {\em a priori} what the pitch-angle scattering rate should be in a given astrophysical or heliospheric context, since it depends on kinetic processes that excite very high frequency fluctuations (relative to the MHD-scale turbulence that is better understood).  We return to this point below when discussing the implications of our results for the solar wind.   

We find that the physics of particle energization by MHD turbulence depends on the ratio of the mean free path set by pitch-angle scattering ($\lambda_{\rm mfp} \simeq v/\nu$) to the scale of the turbulent fluctuations $L$.   For low scattering rates, i.e., $\lambda_{\rm mfp} \gtrsim L$, the particle energization is analogous to that in the  collisionless limit and occurs primarily through quasi-resonant interactions with turbulent fluctuations in the local strength of the magnetic field.   In linear theory, these field compressions would be associated with slow ($\beta \gg 1$) or fast ($\beta \ll 1$) magnetosonic modes and the resulting wave-particle interaction is often called transit-time damping (TTD).   Pitch angle scattering increases the efficiency of this quasi-resonant heating mechanism by reducing the parallel acceleration of particles out of resonance and converting $v_\parallel$ heating into $v_\perp$ heating, thus increasing the magnitude of the $\mu \nabla B$ force associated with magnetic field compressions \citep{Achterberg1981}.   The net heating rate for thermal particles can be a factor of $\sim 3-10$ larger in the presence of modest pitch-angle scattering  ($\nu \sim c_s/L$;  Fig. \ref{fig:nusHeating}).

In the limit of rapid pitch-angle scattering, the physical process leading to particle energization is fundamentally different.  Our numerical results in this limit are consistent with the analytic calculations of \citet{Ptuskin1988a} and \citet{Chandran2004}.  Physically, at high scattering rates, particles are nearly pinned to the frame of the scatterers.   To first approximation, particles thus adiabatically heat and cool as they move through density fluctuations in the plasma.  Spatial diffusion of particles relative to the density fluctuations ultimately makes this energy change diffusive rather than reversible.   Note that the  energy source for  particle energization in the rapid scattering limit is  turbulent density compressions and rarefactions, rather than the turbulent magnetic field compressions that are important at low scattering rates.    In linear theory, these density fluctuations are associated with fast modes at $\beta \gg 1$ and both slow and fast modes at $\beta \ll 1$.   In addition to coupling to the turbulent density compressions, if the scatterers are moving relative to the fluid frame, pitch-angle scattering produces a Fermi-like velocity kick that corresponds to a velocity diffusion coefficient that is independent of velocity for non-relativistic particles (eq. \ref{eq:moving_scat} and Fig. \ref{fig:D_moving}).

One of our primary motivations for including the effects of pitch-angle scattering was to assess whether pitch-angle scattering increases the efficiency of particle {\em acceleration} by MHD turbulence, i.e., the generation of a non-thermal tail in the distribution function.    We find that this is not the case at low scattering rates where quasi-resonant interactions with magnetic field compressions dominate the velocity diffusion (see Fig. \ref{fig:thermalDist}).   Although particle heating is quite effective, with heating rates for thermal protons comparable to the energy cascade rate in the turbulence, the distribution function does not develop a significant non-thermal tail.   This conclusion from our test particle simulations is consistent with solutions of the Fokker-Planck equation given the velocity diffusion coefficient we find at low scattering rates, namely $D \propto v$ for $v \gtrsim c_s$ (Fig. \ref{fig:D_lowNu}).

For high scattering rates, $\nu \sim 10^{2-4} c_s/L$, we find that the distribution function can develop a non-thermal tail, but only under certain relatively restrictive conditions.   The power-law tail arises due to the strong scaling $D \propto v^4$ that occurs at intermediate velocities $\sim {\rm few} \, c_s$ in the rapid scattering limit (eq. \ref{eq:ptuskin_D_B} and Fig. \ref{fig:D_highNu}).   At high particle velocities, the diffusion coefficient varies much more weakly with velocity ($D \propto v$) and so does not produce efficient acceleration.    Physically, the transition between $D \propto v^4$ and $D \propto v$ occurs when the spatial diffusion time ($\propto \nu/v^2$) across an eddy becomes shorter than the  eddy turnover time, as predicted in \citet{Ptuskin1988a}'s original analysis.   This produces a significant power-law tail only if the transition between the two diffusion regimes occurs at a velocity comparable to the thermal velocity of the particles of interest (Fig. \ref{fig:powerLawTails}).   This requires fine tuning between the pitch-angle scattering rate and the thermal velocity of the particles.

What are the implications of our results for particle heating and
acceleration in the solar wind, the most well-studied and well-constrained
astrophysical plasma to which our results can be applied?    This depends to some extent
on the effective scattering rate in the low-collisionality solar wind.
Given the ubiquitous {\rm in situ} evidence for significant
temperature anisotropies in the thermal plasma in the solar wind (e.g.,
\citealt{Kasper2002}) it is likely that the effective scattering rate for the thermal plasma
cannot be much larger than the expansion rate of the solar wind $\sim
R/v_{\rm wind}$  (where $R$ is the local radius, i.e., $\sim 1$ AU near
Earth).  On the other hand, the measured temperature anisotropies in
the solar wind also appear to be constrained by the thresholds for
mirror, firehose, and (to a lesser extent) ion cyclotron instabilities,
suggesting  that there is some minimum level of pitch-angle scattering
even in  nearly collisionless epochs.   Expansion of the solar wind
inevitably drives solar wind plasma towards these instability
thresholds (e.g., \citealt{Hellinger2008}).   Together, these
arguments suggest that the effective scattering rate for thermal plasma is $\nu \sim
v_{\rm wind}/R$, which is in turn $\sim c_s/L$ where $L \sim 0.03 R$
is the outer-scale of turbulence in the near-Earth solar wind (e.g.,
\citealt{Howes2008}) and $v_{\rm wind} \sim 10 c_s$.   Measurements of anisotropies in the distribution function at $v \sim 10 c_s$ (at distances of a few AU) suggest that the mean free path is of order an AU \citep{Gloeckler1995}, implying a comparable effective scattering rate for super-thermal particles.    These estimates
are probably only accurate to  a factor of $10$, but this is
sufficient to draw several important conclusions.  

First, pitch-angle scattering likely enhances  heating of the thermal particles by the
compressible part of the cascade  (which is primarily slow modes in
the near-Earth solar wind; \citealt{Yao2011,Howes2012}) by factors of a few-10
(Fig. \ref{fig:nusHeating}).     Secondly, Figure \ref{fig:thermalDist} shows that even in the
presence of pitch-angle scattering with $\nu \sim c_s/L$, the particle
distribution function does not develop a significant non-thermal tail
due to  MHD-scale turbulent fluctuations.   Moreover, even if we optimistically assume that $\nu \gtrsim 100 \, c_s/L$ (in spite of the  evidence for anisotropic distribution functions in the solar wind), our results do {\em not} imply that rapid pitch-angle scattering would produce a power-law tail in the proton distribution function in the near-Earth solar wind.   The reason is that the particle acceleration time implied by Figure \ref{fig:D_highNu} ($t_{\rm a} \simeq v^2/D$) is longer than the expansion time of the solar wind $R/v_{\rm wind} \sim L/c_s$ over the entire velocity regime in which $D \propto v^4$.   This is even more true when we take into account that the compressional energy in the solar wind -- which is the important energy source in the presence of pitch-angle scattering -- is somewhat smaller than in our simulations:   e.g., $\delta n/n \simeq 0.1$ in the solar wind vs. $\delta n/n \simeq 0.23$ in our fiducial $\beta = 1$ simulation (see \S \ref{sec:turbulenceProperties}).    Thus in the context of the near-Earth solar wind, adiabatic cooling (not present in our calculations) dominates particle acceleration by MHD turbulence even for very rapid pitch-angle scattering; this would preclude the power-law tail seen in Figure \ref{fig:powerLawTails} from developing. 

Taken together, these considerations strongly argue against the MHD
scale fluctuations as the source of the $v^{-5}$ tail in the proton
distribution function observed throughout the heliosphere
\citep{Gloeckler2003,Fisk2006}.   It does not, of course, preclude
that kinetic-scale physics acting on small scales in the turbulent
cascade could produce such a non-thermal tail (e.g., via reconnection or kinetic Alfv\'en wave turbulence).  In addition, our conclusions are based primarily on the well-studied plasma conditions near $\sim 1$ AU.   If the pitch-angle scattering rate and/or the compressibility of the turbulence are significantly larger at larger radii in the solar wind,  the compressible part of the turbulence may be important for producing a significant non-thermal tail.\footnote{There is some ambiguity in the literature on the radial evolution of the turbulence in the solar wind.   Based on Ulysses data, \citet{Breech2009} quote a factor of $\sim 3$ decrease in $\delta v$ from $\sim 1-10$ AU, which suggests that the turbulence is as or more compressible at large radii as at 1 AU.    However, using Voyager 2 data \citet{Ng2010} find that $\delta v$ decreases by a factor of $\sim 16$ over the same radial range,  suggesting significantly lower compressibility  at large radii.   By contrast, direct measurements of the density fluctuations using Voyager 2 data suggest that $\delta n/n$ {\em increases} with increasing heliospheric radius, albeit with somewhat large error bars \citep{Bellamy2005}.}

Although we have focused the discussion of our results on the solar wind application because of the wealth of {\em in situ} measurements, the calculations presented here are much more broadly applicable.   For example, particle heating and acceleration by MHD turbulence has been widely used to model solar flares and the ``re-acceleration" of cosmic rays in the Milky Way and clusters of galaxies.  Some of these applications would require generalizing our techniques to relativistic particles, but this is straightforward.

\begin{acknowledgements}
We thank Phil Isenberg, Marty Lee, and Anatoly Spitkovsky for useful conversations.   
This material is based on work supported by the National Science
Foundation Graduate Research Fellowship under Grant
No. DGE-1106400. Additionally, this work was supported in part by
NSF-DOE grant PHY-0812811, NASA HTP grant NNX11AJ37G, and NSF grant 
ATM-0752503.   EQ was supported in part by a Simons Investigator award from the Simons Foundation, the David and Lucile Packard Foundation, and the Thomas Alison Schneider Chair in Physics.
Computing time was provided by the National Science Foundation TeraGrid/XSEDE
resource on the Frost, Trestles, and Kraken supercomputers. 
\end{acknowledgements}

\bibliographystyle{apj}
\bibliography{library,cites2}

\begin{thebibliography}{48}
\expandafter\ifx\csname natexlab\endcsname\relax\def\natexlab#1{#1}\fi

\bibitem[{Achterberg(1981)}]{Achterberg1981}
Achterberg, A. 1981, Astronomy \& Astrophysics, 97, 259

\bibitem[{Bale {et~al.}(2009)Bale, Kasper, Howes, Quataert, Salem, \&
  Sundkvist}]{Bale2009a}
Bale, S., Kasper, J., Howes, G., Quataert, E., Salem, C., \& Sundkvist, D.
  2009, Physical Review Letters, 103, 211101

\bibitem[{{Bellamy} {et~al.}(2005){Bellamy}, {Cairns}, \&
  {Smith}}]{Bellamy2005}
{Bellamy}, B.~R., {Cairns}, I.~H., \& {Smith}, C.~W. 2005, Journal of
  Geophysical Research (Space Physics), 110, 10104

\bibitem[{Bieber {et~al.}(1994)Bieber, Matthaeus, Smith, Wanner, Kallenrode, \&
  Wibberenz}]{Bieber1994a}
Bieber, J.~W., Matthaeus, W.~H., Smith, C.~W., Wanner, W., Kallenrode, M.-B.,
  \& Wibberenz, G. 1994, The Astrophysical Journal, 420, 294

\bibitem[{{Bochsler}(2007)}]{Bochsler2007}
{Bochsler}, P. 2007, \aapr, 14, 1

\bibitem[{Boris(1970)}]{Boris1970}
Boris, J. 1970, in {Proceedings of the Fourth Conference on Numerical
  Simulations of Plasmas}, {Naval Research Lab}, 3--67

\bibitem[{{Breech} {et~al.}(2009){Breech}, {Matthaeus}, {Cranmer}, {Kasper}, \&
  {Oughton}}]{Breech2009}
{Breech}, B., {Matthaeus}, W.~H., {Cranmer}, S.~R., {Kasper}, J.~C., \&
  {Oughton}, S. 2009, Journal of Geophysical Research (Space Physics), 114,
  9103

\bibitem[{Chandran(2003)}]{Chandran2003a}
Chandran, B. D.~G. 2003, The Astrophysical Journal, 599, 1426

\bibitem[{Chandran {et~al.}(2010)Chandran, Li, Rogers, Quataert, \&
  Germaschewski}]{Chandran2010}
Chandran, B. D.~G., Li, B., Rogers, B., Quataert, E., \& Germaschewski, K.
  2010, The Astrophysical Journal, 720, 503

\bibitem[{Chandran \& Maron(2004)}]{Chandran2004}
Chandran, B. D.~G., \& Maron, J.~L. 2004, The Astrophysical Journal, 603, 23

\bibitem[{Cho \& Lazarian(2003)}]{Cho2003}
Cho, J., \& Lazarian, A. 2003, Monthly Notices of the Royal Astronomical
  Society, 345, 325

\bibitem[{Fermi(1949)}]{Fermi1949a}
Fermi, E. 1949, Physical Review, 75, 1169

\bibitem[{Fisk \& Gloeckler(2006)}]{Fisk2006}
Fisk, L., \& Gloeckler, G. 2006, The Astrophysical Journal Letters, 640, 79

\bibitem[{Fisk \& Gloeckler(2008)}]{Fisk2008}
---. 2008, The Astrophysical Journal, 686, 1466

\bibitem[{Giacalone \& Jokipii(1994)}]{Giacalone1994}
Giacalone, J., \& Jokipii, J.~R. 1994, The Astrophysical Journal, 430, L137

\bibitem[{Gloeckler(2003)}]{Gloeckler2003}
Gloeckler, G. 2003, AIP Conference Proceedings, 679, 583

\bibitem[{{Gloeckler} {et~al.}(1995){Gloeckler}, {Schwadron}, {Fisk}, \&
  {Geiss}}]{Gloeckler1995}
{Gloeckler}, G., {Schwadron}, N.~A., {Fisk}, L.~A., \& {Geiss}, J. 1995, \grl,
  22, 2665

\bibitem[{Goldreich \& Sridhar(1995)}]{Goldreich1995}
Goldreich, P., \& Sridhar, S. 1995, The Astrophysical Journal, 438, 763

\bibitem[{Gruzinov \& Quataert(1999)}]{Gruzinov1999b}
Gruzinov, A., \& Quataert, E. 1999, The Astrophysical Journal, 520, 849

\bibitem[{Hellinger \& Tr\'{a}vn\'{\i}\v{c}ek(2008)}]{Hellinger2008}
Hellinger, P., \& Tr\'{a}vn\'{\i}\v{c}ek, P.~M. 2008, Journal of Geophysical
  Research, 113, A10109

\bibitem[{Hockney \& Eastwood(1981)}]{Hockney1981}
Hockney, R., \& Eastwood, J. 1981, {Computer Simulation Using Particles} (CRC
  Press)

\bibitem[{Howes {et~al.}(2012)Howes, Bale, Klein, Chen, Salem, \&
  TenBarge}]{Howes2012}
Howes, G.~G., Bale, S.~D., Klein, K.~G., Chen, C. H.~K., Salem, C.~S., \&
  TenBarge, J.~M. 2012, The Astrophysical Journal, 753, L19

\bibitem[{Howes {et~al.}(2008)Howes, Cowley, Dorland, Hammett, Quataert, \&
  Schekochihin}]{Howes2008}
Howes, G.~G., Cowley, S.~C., Dorland, W., Hammett, G.~W., Quataert, E., \&
  Schekochihin, A.~A. 2008, Journal of Geophysical Research, 113, 1

\bibitem[{Jokipii(1966)}]{Jokipii1966}
Jokipii, J.~R. 1966, The Astrophysical Journal, 146, 480

\bibitem[{Jokipii \& Lee(2010)}]{Jokipii2010}
Jokipii, J.~R., \& Lee, M.~A. 2010, The Astrophysical Journal, 713, 475

\bibitem[{Kasper {et~al.}(2002)Kasper, Lazarus, \& Gary}]{Kasper2002}
Kasper, J.~C., Lazarus, A.~J., \& Gary, S.~P. 2002, Geophysical Research
  Letters, 29, 1839

\bibitem[{Kennel \& Engelmann(1966)}]{Kennel1966}
Kennel, C., \& Engelmann, F. 1966, Physics of Fluids, 9, 2377

\bibitem[{Kowal \& Lazarian(2010)}]{Kowal2010a}
Kowal, G., \& Lazarian, a. 2010, The Astrophysical Journal, 720, 742

\bibitem[{Lehe {et~al.}(2009)Lehe, Parrish, \& Quataert}]{Lehe2009}
Lehe, R., Parrish, I.~J., \& Quataert, E. 2009, The Astrophysical Journal, 707,
  404

\bibitem[{Longair(1992)}]{Longair1992}
Longair, M.~S. 1992, {High Energy Astrophysics}, 2nd edn. (Cambridge University
  Press)

\bibitem[{Lynn {et~al.}(2012)Lynn, Parrish, Quataert, \& Chandran}]{Lynn2012}
Lynn, J.~W., Parrish, I.~J., Quataert, E., \& Chandran, B. D.~G. 2012, The
  Astrophysical Journal, 758, 78

\bibitem[{Mace {et~al.}(2000)Mace, Matthaeus, \& Bieber}]{Mace2000}
Mace, R.~L., Matthaeus, W.~H., \& Bieber, J.~W. 2000, The Astrophysical
  Journal, 538, 192

\bibitem[{Maechler {et~al.}(2012)}]{sfsmisc}
Maechler, M., {et~al.} 2012, |sfsmisc|: Utilities from Seminar fuer Statistik
  ETH Zurich, {R} Package Version 1.0-20

\bibitem[{{Ng} {et~al.}(2010){Ng}, {Bhattacharjee}, {Munsi}, {Isenberg}, \&
  {Smith}}]{Ng2010}
{Ng}, C.~S., {Bhattacharjee}, A., {Munsi}, D., {Isenberg}, P.~A., \& {Smith},
  C.~W. 2010, Journal of Geophysical Research (Space Physics), 115, 2101

\bibitem[{Ptuskin(1988)}]{Ptuskin1988a}
Ptuskin, V. 1988, Soviet Astronomy Letters, 14, 255

\bibitem[{Qin \& Shalchi(2012)}]{Qin2012}
Qin, G., \& Shalchi, A. 2012, Advances in Space Research, 49, 1643

\bibitem[{{R Core Team}(2012)}]{RLang}
{R Core Team}. 2012, R: A Language and Environment for Statistical Computing, R
  Foundation for Statistical Computing, Vienna, Austria, {ISBN} 3-900051-07-0

\bibitem[{Schekochihin {et~al.}(2008)Schekochihin, Cowley, Kulsrud, Rosin, \&
  Heinemann}]{Schekochihin2008a}
Schekochihin, A., Cowley, S., Kulsrud, R., Rosin, M., \& Heinemann, T. 2008,
  Physical Review Letters, 100, 081301

\bibitem[{Schekochihin {et~al.}(2009)Schekochihin, Cowley, Dorland, Hammett,
  Howes, Quataert, \& Tatsuno}]{Schekochihin2009}
Schekochihin, A.~A., Cowley, S.~C., Dorland, W., Hammett, G.~W., Howes, G.~G.,
  Quataert, E., \& Tatsuno, T. 2009, The Astrophysical Journal Supplement
  Series, 182, 310

\bibitem[{Shalchi {et~al.}(2004)Shalchi, Bieber, Matthaeus, \&
  Qin}]{Shalchi2004}
Shalchi, A., Bieber, J.~W., Matthaeus, W.~H., \& Qin, G. 2004, The
  Astrophysical Journal, 616, 617

\bibitem[{Sharma {et~al.}(2007)Sharma, Quataert, Hammett, \&
  Stone}]{Sharma2007}
Sharma, P., Quataert, E., Hammett, G.~W., \& Stone, J.~M. 2007, The
  Astrophysical Journal, 667, 714

\bibitem[{Skilling(1971)}]{Skilling1971}
Skilling, J. 1971, The Astrophysical Journal, 170, 265

\bibitem[{Skilling(1975)}]{Skilling1975}
---. 1975, Monthly Notices of the Royal Astronomical Society

\bibitem[{Stone {et~al.}(2008)Stone, Gardiner, Teuben, Hawley, \&
  Simon}]{Stone2008}
Stone, J., Gardiner, T.~A., Teuben, P., Hawley, J.~F., \& Simon, J.~B. 2008,
  The Astrophysical Journal Supplement Series, 178, 137

\bibitem[{Tautz {et~al.}(2013)Tautz, Lerche, \& Kruse}]{Tautz2013}
Tautz, R.~C., Lerche, I., \& Kruse, F. 2013, Astronomy \& Astrophysics, 555,
  A101

\bibitem[{{Tu} \& {Marsch}(1995)}]{Tu1995}
{Tu}, C.-Y., \& {Marsch}, E. 1995, \ssr, 73, 1

\bibitem[{Yan \& Lazarian(2008)}]{Yan2008a}
Yan, H., \& Lazarian, A. 2008, The Astrophysical Journal, 673, 942

\bibitem[{{Yao} {et~al.}(2011){Yao}, {He}, {Marsch}, {Tu}, {Pedersen},
  {R{\`e}me}, \& {Trotignon}}]{Yao2011}
{Yao}, S., {He}, J.-S., {Marsch}, E., {Tu}, C.-Y., {Pedersen}, A., {R{\`e}me},
  H., \& {Trotignon}, J.~G. 2011, \apj, 728, 146

\end{thebibliography}

\end{document}